\documentclass{mem}
\usepackage{natbib}\usepackage{txfonts}\usepackage{balance}
\usepackage{graphicx}
\usepackage[a4paper]{hyperref}
\idline{75}{282}
\begin{document}
\def\teff{$T\rm_{eff }$}
\def\kms{$\mathrm {km s}^{-1}$}

\title{
New views on bar pattern speeds from the NUGA survey
}

   \subtitle{}

\author{
V.\,Casasola\inst{1}, S.\,Garc\'ia-Burillo\inst{2},
F.\,Combes\inst{3}, L.~K.\,Hunt\inst{1}, M.\,Krips\inst{4},
E.\,Schinnerer\inst{5}, A.~J.\,Baker\inst{6}, F.\,Boone\inst{3},
A.\,Eckart\inst{7}, S.\,L\'eon \inst{8}, R.\,Neri\inst{9} 
\and L.~J.\,Tacconi\inst{10}
}

  \offprints{V. Casasola}

\institute{
INAF/IRA-FI, Largo Enrico Fermi 5, 50125 Firenze, Italy \\
\email{casasola@arcetri.astro.it}
\and
Observatorio de Madrid, Alfonso XII, 3, 28014-Madrid, 
Spain
\and
Observatoire de Paris-LERMA, 61 Av. de l'Observatoire, 75014 Paris, France
\and
Harvard-Smithsonian CfA, 60 Garden Street, MS 78 Cambridge, MA 02138
\and
MPIA, K\"{o}nigstuhl 17, 69117 Heidelberg, Germany
\and
Rutgers, the State University of NJ
136 Frelinghuysen Road Piscataway, NJ 08854-8019 
\and
I. Physikalisches Institut, Un. zu K\"{o}ln, Z\"{u}lpicherstrasse 77, 
50937 K\"{o}ln, Germany
\and
IRAM, Avenida Divina Pastora 7, N\'ucleo Central, 18012 Granada, Spain
\and
IRAM, 300 rue de la Piscine, 38406 St Martin d’H\`eres Cedex, France
\and
MPE, Giessenbachstrasse 1, 85741 Garching, Germany
}

\authorrunning{Casasola}

\titlerunning{NUGA survey}

\abstract{
We present a review on bar pattern speeds 
from preliminary results in the context 
of the Nuclei of Galaxies (NUGA) project.
The large variety of molecular circumnuclear morphologies 
found in NUGA is a challenging result that implies 
the refinement of current dynamical models of galaxies.

\keywords{Galaxies: spiral --
Galaxies: active -- Galaxies: nuclei -- 
Galaxies: kinematics and dynamics
}
}
\maketitle{}

\begin{figure}[!h]
\resizebox{\hsize}{!}{\includegraphics[clip=true,angle=-90]{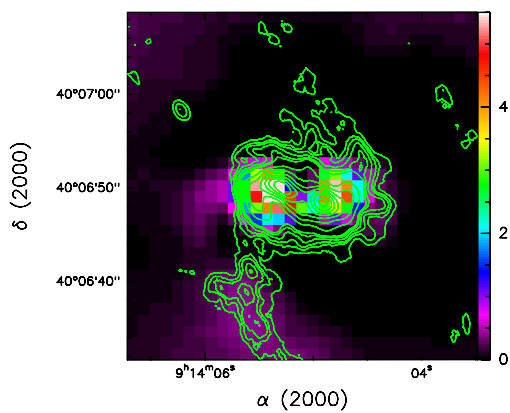}}
\caption{
\footnotesize
$^{12}$CO(1-0) total intensity contours (in green) overlaid on the \textit{residuals} 
(in MJy/sr) of the bulge-disk decomposition performed on the 
3.6 $\mu$m \textit{Spitzer}/IRAC for NGC\,2782 \citep[][]{leslie08}.
}
\label{n2782}
\end{figure}

\section{Introduction}
Accretion of gas onto supermassive black holes (SMBHs) in galactic centers is the source 
of active galactic nuclei (AGN) activity.
AGN must be fed with material coming from the disk of the host galaxy, far away from the 
influence of the BH, and this implies that the gas supply should lose its angular 
momentum during the fueling process.
Different mechanisms for removing the gas angular momentum have been 
proposed, 
% \citep[e.g.][]{jogee06},
processes which can be accomplished 
through non-asymmetric perturbations. 
Gravitational torques due to density waves (e.g. spirals, bars, and lopsidedness perturbations) 
or galaxy-galaxy interactions (e.g. galaxy collisions, mergers, close encounters, and mass accretion)
are capable of removing angular momentum of the rotating gas.
The study of molecular gas, the predominant phase of the interstellar medium in the
nuclei of spiral galaxies, is fundamental to investigate in detail
how AGN activity is related to the molecular gas morphology and kinematics.
The connection between AGN and molecular gas is also complicated by the presence of different scales, 
both spatial and temporal. 
The AGN duty cycle ($\sim$10$^{5-6}$\,years) should be shorter than the lifetime of the feeding 
mechanism \citep[e.g.][]{wada04}, and at present the critical spatial scales for AGN feeding 
($<$10-100\,pc) can only be achieved with mm-interferometers in nearby low-luminosity AGN (LLAGN).
% Viscosity can also play a not negligible role in the fueling process.
% Viscous torques in combination with gravitational torques can produce
% recurrent episodes of activity during the typical lifetime of any
% galaxy \citep{santi05}. 
%In some cases, the efficiency of viscosity 
%may be comparable to that of gravity torques \citep[e.g., NGC\,4579][]{santi08}.

\begin{table*}[!htpb]
\caption{NUGA galaxies studied in detail.}
\begin{center}
\begin{tabular}{lllcl}
\hline
\hline
Galaxy & SB/GB/OD$^{*}$  & SR/GR$^{*}$ & $\Omega_p$
               & Predicted resonances$^{*}$ \\
               &           &       & km s$^{-1}$ kpc$^{-1}$  &  \\
\hline
               &           & \textbf{Galaxies with inflow}   & &  \\
\textbf{NGC\,6574} & NSB $r=0.6$\,kpc                & & & \\
& PSGB $r=1.6$\,kpc & OGR $r=1.6$\,kpc & 65 & UHR $r=1.6$\,kpc \\
&                           & IGR $r=130$\,pc  & & iILR $r=30-900$\,pc  \\
&                         &                            & & oILR $r=1$\,kpc      \\
&                         &                            & & CR $r=2.8$\,kpc      \\
&                         &                            & &  \\
\textbf{NGC\,2782} & NSGB$^{DB}$ $r=1$\,kpc & IGR $r=300$\,pc  & 270 
& ILR $r=200-500$\,pc   \\
&   &    &  & CR $r=1.1$\,kpc                           \\
& PSB$^{DB}$ $r=2.5$\,kpc & OGR $r=1$\,kpc & 65  & ILR $r=1.3$\,kpc \\
&  &    &   & UHR $r=3.6-3.8$\,kpc \\
& & &     & CR $r=5$\,kpc        \\
&  &                &       &  \\
\textbf{NGC\,3147} & PSB/OD $r=1-1.5$\,kpc  & IGR $r=2$\,kpc
& 125 & UHR $r=2$\,kpc       \\
&             &                  &     & CR $r=3$kpc          \\
&             & OGR $r=4$\,kpc &     &                        \\
&             &                        &     &  \\
\textbf{NGC\,4579} & NSB/OD$^{DB}$ $r=200$\,pc & ISB $r=200$\,pc
& 270 & UHR $r=200$\,pc      \\
&             &                &       & CR $r=1$\,kpc           \\
& PSB$^{DB}$ $r=6$\,kpc &  & 50  & iILR $r=500$\,pc     \\
&             & OGR $r=1$\,kpc &     & oILR $r=1.3$\,kpc \\
&                          &                &     & UHR $r=3.8$\,kpc     \\
&                          &                &     & CR $r=6$\,kpc        \\
&                          &                &       &  \\
\hline
&           & \textbf{Galaxies without inflow}      &   &  \\
\textbf{NGC\,4826}  & {\scriptsize $m=1$ fast trailing spiral waves} & 
& 1500 & CR $r=0.1$\,kpc  \\
&              &               &   & OLR $r=0.35$\,kpc \\
&              &               &   &  \\
\textbf{NGC\,7217} & OD $r=3.7$\,kpc & NSGR $r=0.8$\,kpc & 
80 & ILR $r=1$\,kpc \\
&          & ISR $r=2.2$\,kpc  & & UHR $r=2.2$\,kpc \\
&          &                           & & CR $r=3.7$\,kpc \\
&          & OSR $r=5.4$\,kpc  & & OLR $r=4-5$\,kpc \\
&          &                           & &  \\
\textbf{NGC\,4569} & PSGB $r=3.3$\,kpc & & 60 & ILR $r=600$\,pc \\
&                          &       &         & CR $r=4$\,kpc \\
\hline
\hline
\end{tabular}
\label{table}
\end{center}
\begin{list}{}{}
   \item[$^{\mathrm{}}$]$^{*}$ These quantities are uncertain to roughly $\pm20\%$,
   \item[$^{\mathrm{}}$]$^{DB}$ Decoupled bars,\\
   ILR = Inner Lindblad Resonance, iILR = inner ILR, oILR = outer ILR,\\
   OLR = Outer Lindblad Resonance, UHR = Ultra-Harmonic Resonance,\\ 
   SB = Stellar bar, GB = Gas bar, NSB = Nuclear stellar bar, NSGB = Nuclear stellar gas bar,\\
   ISB = Inner stellar bar, PSB = Primary stellar bar, PSGB = Primary stellar gas bar,\\
   SR = Stellar ring, NSGR = Nuclear stellar gas ring, ISR = Inner stellar ring, 
   OSR = Outer stellar ring,\\
   GR = Gas ring, IGR = Inner gas ring, OGR = Outer gas ring,\\
   OD = Oval distortion,\\
   CR = corotation.\\ 
   \end{list}
   \end{table*}

The NUclei of GAlaxies (NUGA) project, a large-scale international 
collaboration for a high-resolution 
($\leq$1$^{\prime\prime}$) and high-sensitivity carbon monoxide (CO) survey made with the 
Plateau de Bure IRAM interferometer, has been initiated
to better understand the mechanisms for gas fueling of AGN.
The survey is based on a sample of 12 nearby LLAGN, which span the whole sequence of activity types (Seyferts, LINERs, and transition objects).
NUGA galaxies already analyzed show a ``zoo'' 
of molecular gas morphologies which characterize 
the inner kpc of the galaxies. 
These morphologies include one and two armed instabilities \citep[][]{santi03, krips05}, 
rings and nuclear spirals \citep[][]{francoise04,vivi08}, and large-scale bars and 
two-arm spirals \citep[][]{boone07,lindt08,leslie08,santi08}. 

\begin{figure}[!h]
%\resizebox{\hsize}{!}{\includegraphics[clip=true]{8874fig16.ps}}
\resizebox{\hsize}{!}{\includegraphics[clip=true,angle=-90]{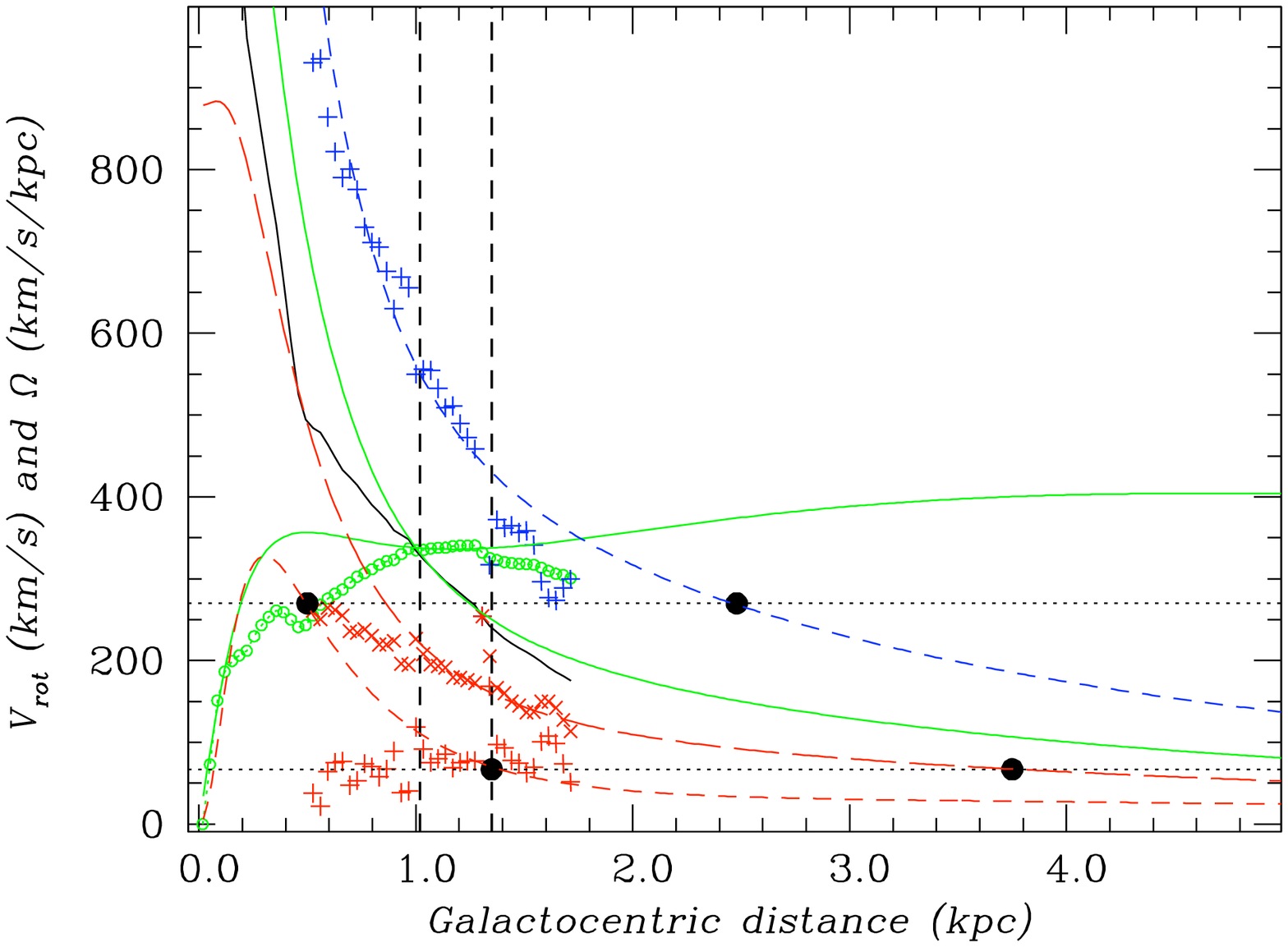}}
\caption{
\footnotesize
Rotation and derived frequency curves $\Omega$, 
$\Omega\pm k/2$ and $\Omega - k/4$, and the empirical 
ones derived from our $^{12}$CO observations for NGC\,2782.
% The (green) open circles represent the empirical V$_{rot}$ 
% coinciding with the empirical circular velocity traced by 
% a solid line.
% The black and green solid lines trace the empirical and 
% theoretical $\Omega$ curves, respectively.
% The long-dashed line shows the $\Omega - k/4$ curve,
% and the short-dashed ones are the $\Omega\pm k/2$ ones.
% Data points are shown as x and +.
% Horizontal dotted lines are the two pattern speeds, and filled circles
% represent the resonances.
The vertical dashed lines indicate the nuclear bar length ($\sim$1\,kpc)
and the ILR of the outer oval ($\sim$1.3\,kpc), close to CR of the nuclear bar
($\sim$1.1\,kpc). For details see \citet{leslie08}.
}
\label{rot-cur}
\end{figure}

\begin{figure}[!h]
\resizebox{\hsize}{!}{\includegraphics[clip=true]{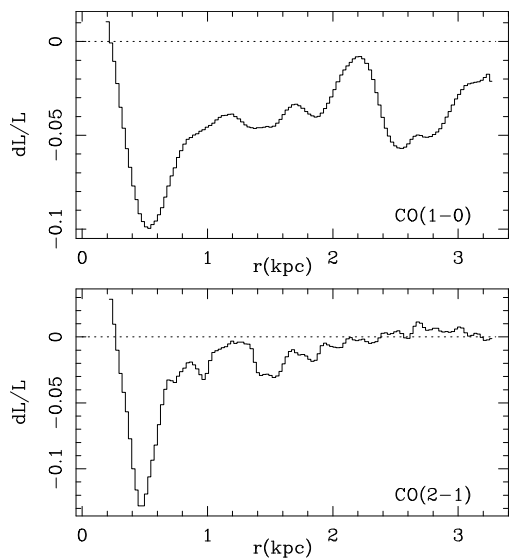}}
\caption{
\footnotesize
The relative torque, $dL/L$, derived for NGC\,2782
in $^{12}$CO(1-0) (\textit{upper panel}) and $^{12}$CO(2-1) 
(\textit{bottom panel}) using the potential obtained from 
the 3.6 $\mu$m \textit{Spitzer}/IRAC image.
}
\label{torques}
\end{figure}

Here, we present the first results regarding bar pattern speeds
obtained with a statistical review of NUGA galaxies already 
analyzed in detail.

\section{NUGA analysis}
\subsection{CO morphologies}
For each NUGA galaxy, we study the overall distribution of the 
molecular gas in the inner kpc, both in $^{12}$CO(1-0) and 
$^{12}$CO(2-1) emission.
Figure \ref{n2782} shows the CO morphology found for 
the starburst/Seyfert 1 NGC\,2782 \citep{leslie08}.
The $^{12}$CO(1-0) emission is aligned with the 
nuclear stellar bar of radius $\sim$1\,kpc.
At the ends of the nuclear bar, the CO changes direction 
and traces diffuse asymmetric spiral arms extending 
to the north and south and aligned with an outer stellar oval 
reminiscent of a weak primary bar.

\subsection{Gravity torques}
The NUGA analysis continues by interpreting 
the distribution and kinematics of the gas, case-by case, 
in terms of feeding mechanisms.
To do this, we compute the gravitational torques 
by first using high-resolution optical and near-infrared 
images of the galaxies to derive gravitational potential.
We can then quantify the efficiency of the stellar 
potential to drain the gas angular momentum: 
negative torques mean that the gas is losing angular 
momentum and flowing into the central regions, 
while positive torques are driving the gas outwards.
Figure \ref{torques} shows the relative torques, $dL/L$,
derived for NGC\,2782 from the
3.6 $\mu$m \textit{Spitzer}/IRAC image.
Both $^{12}$CO(1-0) and
$^{12}$CO(2-1) emission give a similar
picture: \textit{the average torques are negative in the inner few kpc
of NGC\,2782, down to the resolution limit of our images.}

\subsection{Rotation curves}
The final step is to derive the rotation curve 
from our CO observations and to compare it with 
derived frequency curves.
Figure \ref{rot-cur} shows this comparison performed
for NGC\,2782.
In this figure, one can identify the nuclear bar with 
$\Omega_b$$\sim$270\,km\,s$^{-1}$\,kpc$^{-1}$
and corotation (CR) at $\sim$1.1\,kpc, and the primary bar with 
$\Omega_b$$\sim$65\,km\,s$^{-1}$\,kpc$^{-1}$
and an Inner Lindblad Resonance (ILR) at $\sim$1.3\,kpc, 
very close to the CR of the nuclear bar.

\section{Bar pattern speeds and resonances}
By analyzing the gravitational torques on 
the molecular gas, we found net inflow 
in four galaxies studied in detail so far, 
with particularly strong cases in NGC\,2782 
and NGC\,4579.
Table \ref{table} summarizes the first results 
obtained considering seven NUGA galaxies.
\textit{
The common feature shared by NGC\,2782 and NGC\,4579
is two decoupled bars with overlapping resonances: 
one bar is rapidly rotating 
($\Omega_b$$\sim$270\,km\,s$^{-1}$\,kpc$^{-1}$)
and the other rotates more slowly 
($\Omega_b$$\sim$50-65\,km\,s$^{-1}$\,kpc$^{-1}$).}
% The common feature shared by NGC\,2782 and NGC\,4579
% is the \textit{overlap of resonances} in the
% two decoupled bar-like ($m=2$) perturbations.
Such resonances and kinematic decoupling 
are fostered by a large central mass concentration 
(e.g. stellar bulge) and high gas fraction.

\section{Conclusions}
The analysis of the whole NUGA sample will provide a 
clearer picture on the distribution and kinematics 
of molecular gas in the nuclei of AGN and on the 
different mechanisms for gas fueling.
NUGA also represents a pilot study for the investigation 
of fueling mechanisms at high redshift, which will
be possible thanks to instruments of the immediate 
future, such as ALMA.

\begin{acknowledgements}

V. Casasola wish to thank the organizers of Omega08,
an interesting and friendly workshop.
The NUGA team thanks the scientific and technical staff at 
the IRAM for their help during our observations.
\end{acknowledgements}

\bibliographystyle{aa}

\end{document}